\UseRawInputEncoding
\documentclass[twocolumn,showpacs,preprintnumbers,superscriptaddress,amsmath,amssymb]{revtex4-1}

\usepackage[colorlinks=true,bookmarks=false,citecolor=blue,urlcolor=blue]{hyperref}

\usepackage{amsmath,amssymb}
\usepackage{graphicx}
\usepackage{verbatim}
\usepackage{enumitem}
\usepackage{color}
\usepackage[english]{babel}
\usepackage{bm}
\usepackage{natbib}
\usepackage{textcomp}

\usepackage{epstopdf}
\usepackage{setspace}

\begin{document}

\title{FIBER-OPTIC VECTOR ACOUSTIC RECEIVER BASED ON ADAPTIVE HOLOGRAPHIC INTERFEROMETER}

\author{R.V.~Romashko}
 \email{romashko@iacp.dvo.ru}
\affiliation{Institute of Automation and Control Processes, Far Eastern Branch, Russian Academy of Science, Vladivostok 690041, Russia}
\affiliation{Far Eastern Federal University, Vladivostok, Russia}

\author{D.V.~Storozhenko}
\affiliation{Institute of Automation and Control Processes, Far Eastern Branch, Russian Academy of Science, Vladivostok 690041, Russia}

\author{M.N.~Bezruk}
\affiliation{Institute of Automation and Control Processes, Far Eastern Branch, Russian Academy of Science, Vladivostok 690041, Russia}

\author{D.A.~Bobruyko}
\affiliation{Institute of Automation and Control Processes, Far Eastern Branch, Russian Academy of Science, Vladivostok 690041, Russia}

\author{Yu.N.~Kulchin}

\affiliation{Institute of Automation and Control Processes, Far Eastern Branch, Russian Academy of Science, Vladivostok 690041, Russia}
\affiliation{Far Eastern Federal University, Vladivostok, Russia}

\begin{abstract}
A mobile scalar-vector acoustic receiver is proposed, experimentally implemented and investigated. The key components of the receiver are (i) the six-channel fiber-optic coil-type sensor configured as to detect three projections of acoustic intensity vector, (ii) the six-channel optical phase demodulator based on six-channel adaptive holographic interferometer configured with use of dynamic holograms multiplexed in a photorefractive crystal of cadmium telluride and (iii) the signals recording ADC-based system combined with software package for data processing. Field tests of the developed receiver applied for obtaining scalar and vector parameters of acoustic waves generated by a stationary and moving acoustic source in open air and water area are carried out. Experimental results show perceptiveness of use of the fiber-optical adaptive interferometry system for bearing of weak acoustic sources in real conditions.

\end{abstract}

\maketitle

\section{Introduction}
Acoustic monitoring of sea areas, observation of the underwater situation, control of underwater vehicles, including of unmanned-type  and a number of other applied problems require highly sensitive means of registering weak hydroacoustic signals \cite{1,2}. Such problems are usually solved by using both classical hydroacoustic sensors based on electrical transducers (piezoelectric, electrodynamic, capacitive, etc.) and currently developing fiber-optic sensors \cite{3,4}. Using interferometry principles the for fiber-optic sensors signal processing makes it possible to detecting ultra-weak signals due to potentially high sensitivity of optical interferometer \cite{5,6,7,8}. In addition, methods of adaptive interferometry based on two-wave mixing at dynamic hologram recorded in a photorefractive crystal (PRC) make it possible to effectively stabilize the operating point of the optical interferometer without any additional stabilization procedures \cite{9,10,11}. Such technique allows to successfully apply the adaptive interferometers for measuring weak signals under the conditions of influence of external noise factors such as random mechanical impacts, industrial noisy sound and vibrations, temperature or pressure drift, etc. \cite{12,13,14,15,16,17,18}. Adaptive interferometers were being already applied for acoustic and hydroacoustic measurements \cite{19,20,21,22,23}, including use of scalar-vector methods for detection of an acoustic signal \cite{24,25}. In the paper \cite{25}, the adaptive laser scalar-vector hydroacoustic measuring system which provides determination of full vector of acoustic intensity is proposed. However experimental study of this system was carried out in a narrow space (water tank) which dimensions are comparable with acoustic wavelength, and only for one frequency of sound. These conditions cannot exclude distortion of the obtained experimental results caused by influence of sound reflections from the water boundaries. Therefore, the applicability of the proposed system for acoustic bearing requires a confirmation.

In this work, a mobile scalar-vector acoustic receiver based on six fiber-optic coil type acoustic sensors and six-channel adaptive interferometer using dynamic holograms multiplexed in a photorefractive crystal is developed and investigated. The optimized construction of the fiber-optic acoustic sensors has allowed increasing the receiver sensitivity up to 51 V/Pa and reducing the detection threshold down to 0.5 mPa. The performance of the developed receiver has been studied in the field tests on detection and measurement of weak hydroacoustic signals in a marine area and acoustic signals in an open air. The obtained results confirm the possibility of reliable bearing an acoustic source in a real environment which is characterized by industrial vibrations, artificial and natural acoustic noises, wind, sea waves, etc.

\section{The receiver architecture}
The developed acoustic receiver consists of two main blocks: (i) the six-channel fiber-optic sensor which provides primary registration of the acoustic signal, and (ii) the registering system for demodulation, recording and processing of optical signals from the sensor (Fig. 1). The fiber-optic sensor comprises six spatially spaced coil-type sensing elements. The base of each element is fabricated from XPS extruded polystyrene and has cylindrical shape with the diameter of 65 mm and the height of 25 mm. A multimode optical fiber with a core diameter of 62.5 $\mu$m, numerical aperture $NA$ = 0.22 and length of 10 m is wound on the cylinder (30 winds). The impact of acoustic pressure on the coil material causes its stretching deformation which, in turn, leads to phase modulation of the light passing through the optical fiber wound on the cylinder. The six coils are arranged on a metal frame in three pairs in such way that each pair is oriented along one of the three orthogonal axes of space (Fig. 1). This arrangement of the coils of the fiber-optic sensor ensures that all three components of the acoustic pressure gradient are to be measured \cite{25}. The distance between coils for each pair determines not only the overall dimension of the sensor, but also the operating frequency range of the receiver. According to the recommendations of Ref. \cite{26}, the acoustic signals with wavelength in the range 4d$_{0}$\textdiv10d$_{0}$ can be correctly detected and processed by means of a vector receiver. The distance between coils, d0, for the developed sensor is amounted to 25 cm. Consequently, the wavelength range is 1\textdiv2.5 m. It corresponds to the frequency range of 130\textdiv340 Hz for air and 600\textdiv1000 Hz for aquatic environment.

The registering system includes a six-channel adaptive holographic interferometer, a six-channel ADC and PC (Fig. 1). The adaptive interferometer performs phase demodulation of light waves coming from the fiber-optic sensors. The interferometer is based on a photorefractive crystal (CdTe:V) of (100)-cut where six dynamic holograms associated with six measuring channels are continuously recorded in the orthogonal geometry of vector two-wave mixing \cite{25}. The dynamic hologram operates as an adaptive beam combiner and provides stabilization of the interferometer operating point. This allows an interferometer to adapt itself to all low-frequency noises and measure the weak physical quantities (acoustic signals, etc.) in out-of-laboratory conditions which are characterized by accidental mechanical impacts, industrial noises, temperature or pressure drift etc. It is worth noting that the dynamic holograms are recorded in orthogonal geometry of vectorial two-wave mixing where all signal waves from optical fibers are directed to the photorefractive crystal orthogonally to the reference wave \cite{27}. Such geometry provides independent operation of all the measuring channels of the adaptive interferometer, and excludes a crosstalk between the channels \cite{28,29}.

All parts of the registering system are packaged into the portable waterproof box and supplied from the power battery with capacity of 3 A·h which provides autonomous operation of the whole system during 3 hours. The fiber-optic sensor is connected with the adaptive interferometer via optical patch cords with length of 30 m. Due to high coherence of the laser used in the receiver (Nd:YAG laser, $\lambda$ =1064 nm, optical power 1000 mW, SLM, coherence length 300 m), the distance between the fiber-optic sensor and the registration system can be extended up to 100 m if necessary.

\begin{figure}[htbp]
\includegraphics[width=\linewidth]{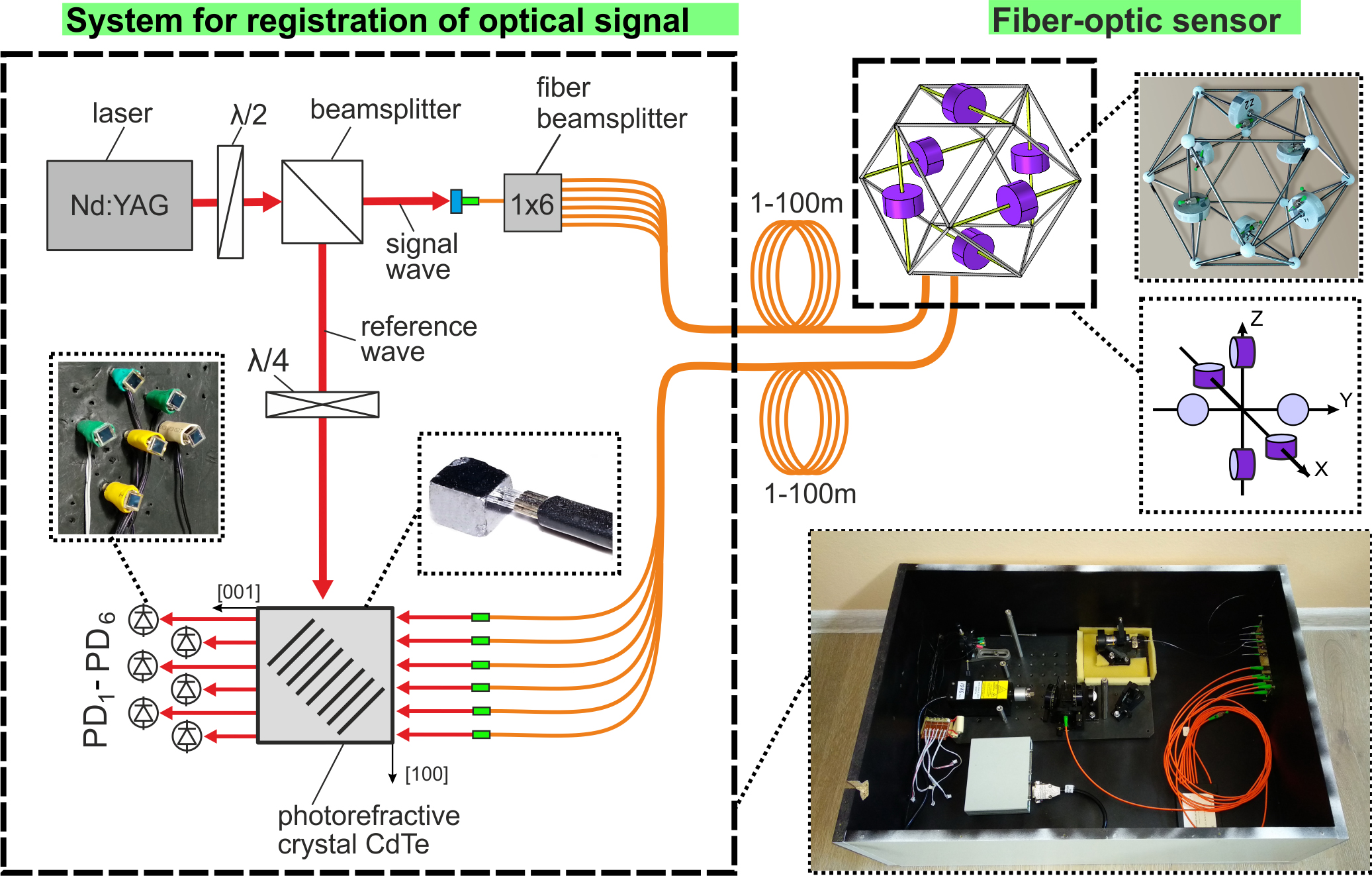}
\caption{Diagram of the mobile fiber-optic scalar-vector acoustic receiver on the base of an adaptive holographic interferometer.}
\end{figure}

The recorded signals from the six photodetectors PD$_{1}$-PD$_{6}$ which measure the intensity of light beams in six interferometer channels are fed, through a DC-blocking filter, to the input of a multichannel amplifier with a controlled gain. Then each signal is digitized by a 12-bit ADC with a sampling rate of 57 kHz per a channel and fed to the processing software. At the first stage of signal processing the algorithm performs signal filtering and Fourier transform. The second stage, the scalar and vector parameters of the acoustic field, in particular the acoustic intensity vector, are calculated for selected frequency components. The mathematical procedure for the calculations is described in details in work \cite{25}. Single measurement and processing cycle takes time 0.2\textdiv0.6 sec.

Here we also applied the two-stage calibration of the developed receiver which was proposed in the paper \cite{25}. Such calibration allows compensating the inevitable difference in the adaptive interferometer channels sensitivity caused by the inhomogeneity of a PRC properties over its volume, different light wave amplitudes at the output from the fiber-optic splitter, difference in the coils geometry, length of wound fiber in a fiber-optic sensing elements, etc.

In contrast to the adaptive hydrophone system presented in the work \cite{23}, the size of fiber-optic sensor coils was increased by 30\%, the length of optical fiber wound on the coils was increased from 5 to 10 m. This results in increasing the sensitivity of the receiver from (6.7 $\pm$ 0.1)rad/Pa or (29 $\pm$ 0.12)V/Pa up to (11.7 $\pm$ 0.2)rad/Pa or (51 $\pm$ 0.19)V/Pa. In addition, the use of a low-noise amplifier made it possible to improve the signal-to-noise ratio (SNR) of the sensor. Such optimization of the system has provided a reduction of the minimum detectable acoustic pressure from 3 mPa down to 0.5 mPa.

\section{Testing the vector receiver on an open air}

The developed vector acoustic receiver was tested in an air environment on an open area. Diagram of the experiment arrangement is shown in Fig.2. The fiber-optic sensor was fixed on a rod at the height of 2 m from the ground. The acoustic source generating a harmonic acoustic signal at the frequency of 340 Hz and sound level 40dB (re 20 µPa) was moving along a circle trajectory with radius of 5 m around the rod located in the center with variable angular velocity (in the range of 5\textdiv10 deg/s). In the experiment, the trajectory plane was set at three different levels of height from the ground: (1) h$_{1}$ = 2 m (at the same height with the sensor), (2) h$_{2}$ = 1 m (below the sensor), and (3) h$_{3}$ = 3 m (above the sensor).

Full turn of the acoustic source along the circle was completed during 1 min. The recorded signal was processed by the software algorithm that calculated the acoustic intensity vector I inside the time interval of 0.6 sec. Thus, for a single pass of the source, a set of 95 values of the vector was obtained. Fig. 2b shows the temporal evolution of orthogonal components of the vector I. As seen, the z-components of the vector I obtained for three height levels (I$_{1z}$, I$_{2z}$, and I$_{3z}$) were kept constant during the movement and correspond to constant heights of the acoustic source. In contrast, the x- and y-components (I$_{1x}$ and I$_{1y}$) were changed in the way close to harmonic dependency which corresponds to the trajectory of the source movement along the circle. It worth to note that deviation I$_{1x,y}$(t) from harmonic functions is related with non-constant speed of the source movement (it accelerates in the time interval of 45\textdiv55 samples and slowdown in the time interval of 65\textdiv85 samples).

\begin{figure}[htbp]
\includegraphics[width=\linewidth]{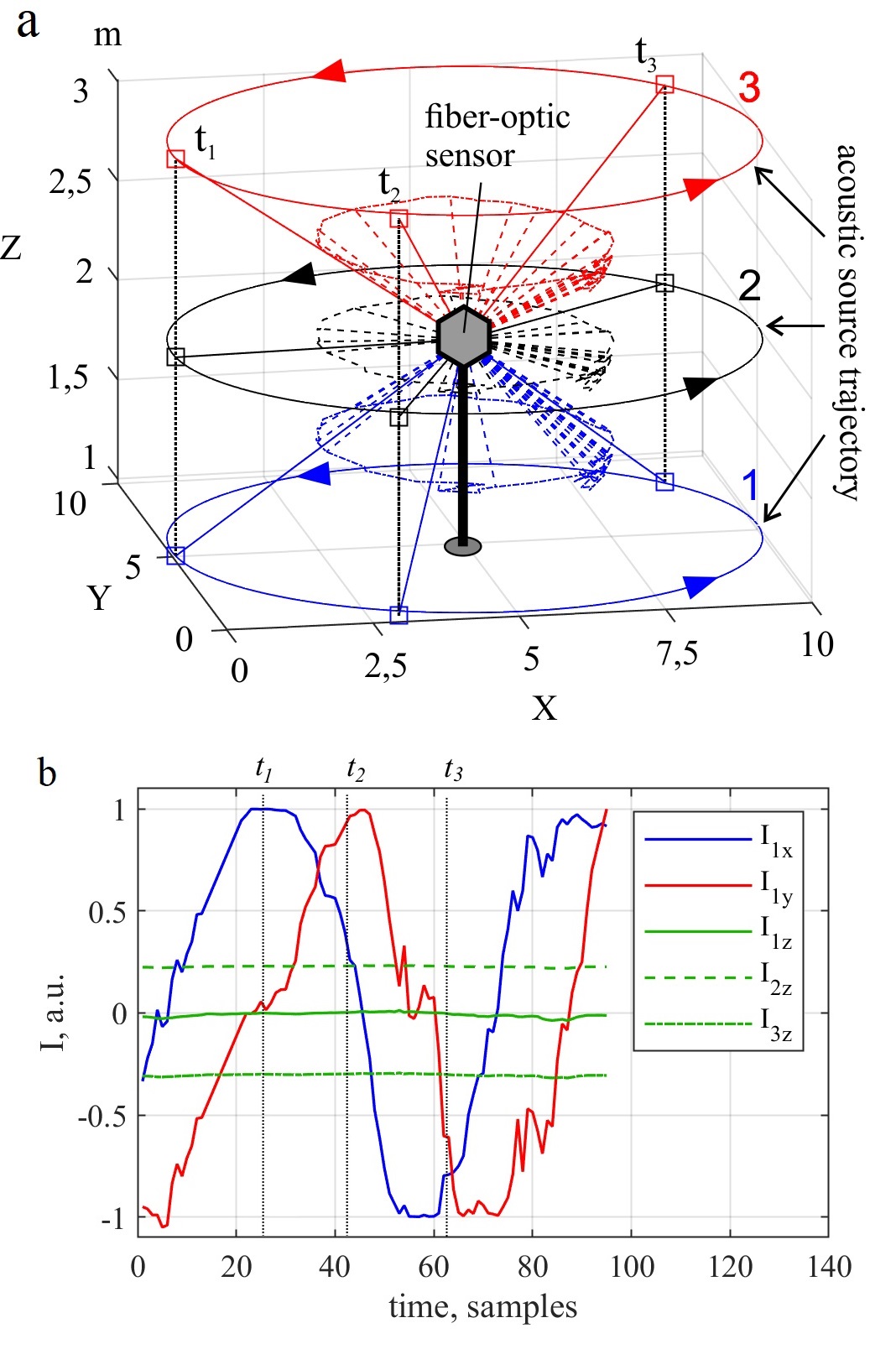}
\caption{Experimental results of testing the fiber-optic scalar-vector receiver on an open air: (a) arrangement of the receiver position and the acoustic source trajectories (1, 2, 3) accompanied by the bearing vectors D (dashed lines) obtained with use of the receiver for three levels of the source trajectory at heights h$_{1}$ (black), h$_{2}$ (blue) and h$_{3}$ (red); (b) temporal evolution of the selected orthogonal components of the acoustic intensity vector. t$_{1}$ = 25th sample, t$_{2}$ = 42th sample, t$_{3}$ = 62th sample.}
\end{figure}

Bearing to the acoustic source is determined by the unit vector inverse to the vector of the acoustic intensity D=-I/|I|. Fig. 2a shows the scheme of the experimental arrangement with the trajectories of the acoustic source and the obtained bearing vectors D directed the to the source positions. In particular, the continuations of the vectors D built at time moments t$_{1}$ = 25, t$_{2}$ = 42 and t$_{3}$ = 62 are traced to the corresponding trajectory of the source motion. As seen, the angular displacement of the source in the interval between t$_{2}$ and t$_{3}$ is twice greater than the displacement between t$_{1}$ and $_{2}$, which is related with increase of the source speed mentioned above. The maximum root-mean-square (RMS) deviation of the azimuthal angle of direction to the source did not exceed 3.5$^o$. Thus, the fiber-optic scalar-vector receiver can provide reliable real-time obtaining of the direction to a signal source from the measured acoustic intensity vector at open air.

\section{Testing the vector receiver in a seawater area}

The field tests of the developed scalar-vector acoustic receiver applied for the problem of obtaining the direction to a moving source of a hydroacoustic signal were carried out in a sea area. The diagram of the experiment is shown in Fig. 3. The acoustic source being submersed into sea water at the depth of 0.5 m was linearly displaced at the constant speed 0.2 m/sec along a straight line perpendicular to the coastline from point A1 to point A2 at the distance of 12 m. The source emits a periodic acoustic signal with the amplitude of 30 mPa at the frequency of 600 Hz and amplitude modulation at the frequency of 200 Hz. The fiber-optic sensor was located stationary at the depth of 6 m, at the distances of 0.5 m from the bottom, 30 m from the coastline, and 4 m from the trajectory of the source movement. The spatial orientation of the receiver was chosen in such a way that the z-axis was perpendicular to the sea bottom, while the x-axis makes the angle of 20$^o$ with the perpendicular to the coastline.

\begin{figure}[htbp]
\includegraphics[width=\linewidth]{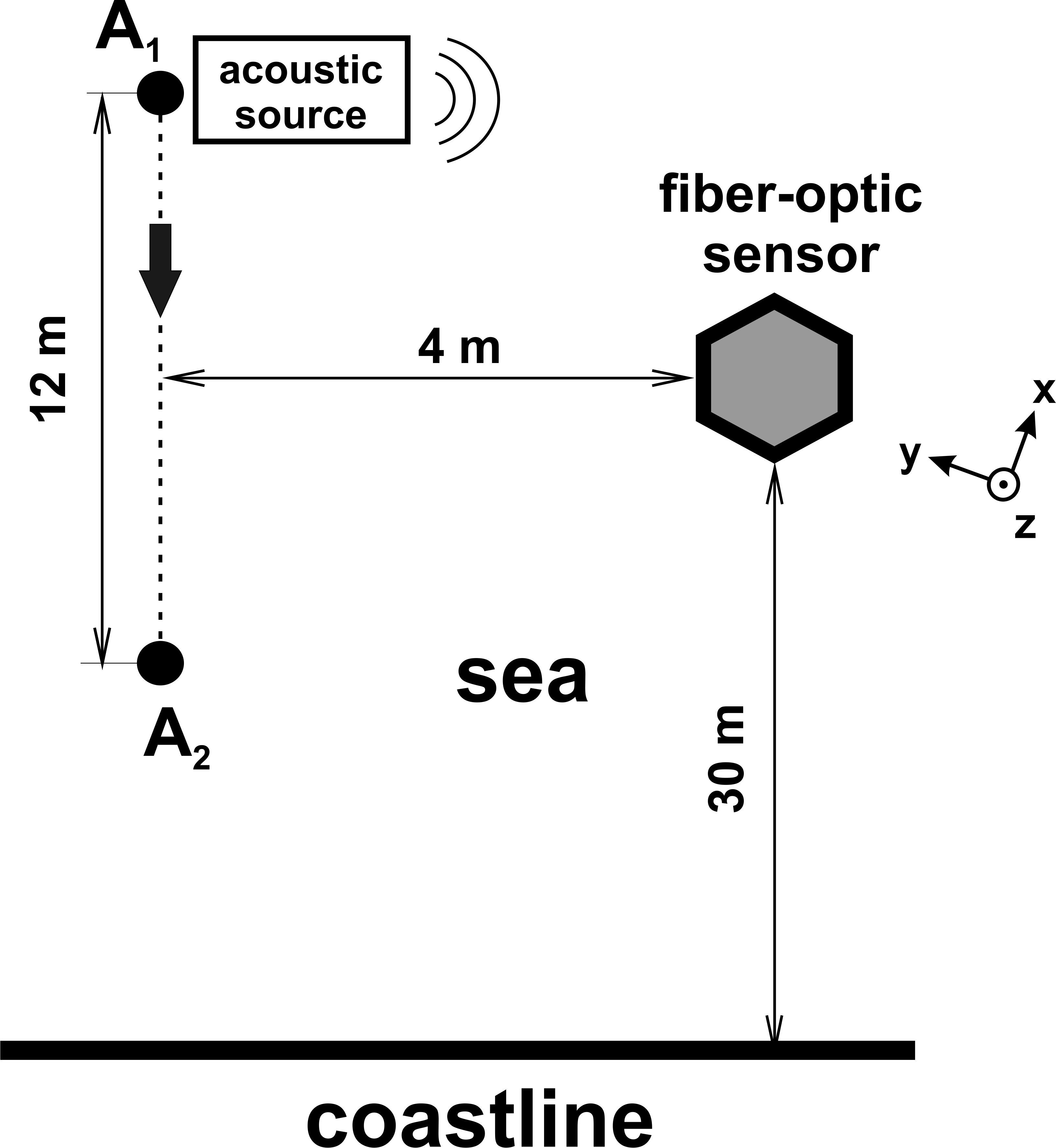}
\caption{Diagram of field test of the fiber-optic vector acoustic receiver in seawater area.}
\end{figure}

The output signal of the receiver was continuously recorded during the full time of the acoustic source movement (~1 min). Fourier spectrum of the received signal is shown in the inset of Fig.4(a). As seen it contains three discrete frequency components – at central, sum and difference frequencies, 600 Hz, 800 Hz and 400 Hz, respectively. For the calculation of the acoustic intensity vector, only two frequency components (600 and 800 Hz) were used, since the rest one (400 Hz) is out of the range which is appropriate for the correct determining of vector parameters (600\textdiv1000 Hz) as was explained above.

Whole measurement time for each discrete frequency component was subdivided into the intervals of 0.5 sec. Fig.4(a) shows the temporal evolution of all orthogonal components of the acoustic intensity vector obtained at frequencies of 600 Hz and 800 Hz. It worth to note that the behavior of the horizontal components of the vector (I$_{x}$ and the I$_{y}$) is similar for both frequencies. Contrary, the vertical component (I$_{z}$) obtained at the frequency of 600 Hz differ from one obtained at the frequency of 800 Hz. This can be caused by sound reflection from the bottom and the surface in condition of shallow sea (the depth of 6.5 m) which affects the acoustic intensity in the z-direction differently at different frequencies.

\begin{figure}[htbp]
\includegraphics[width=\linewidth]{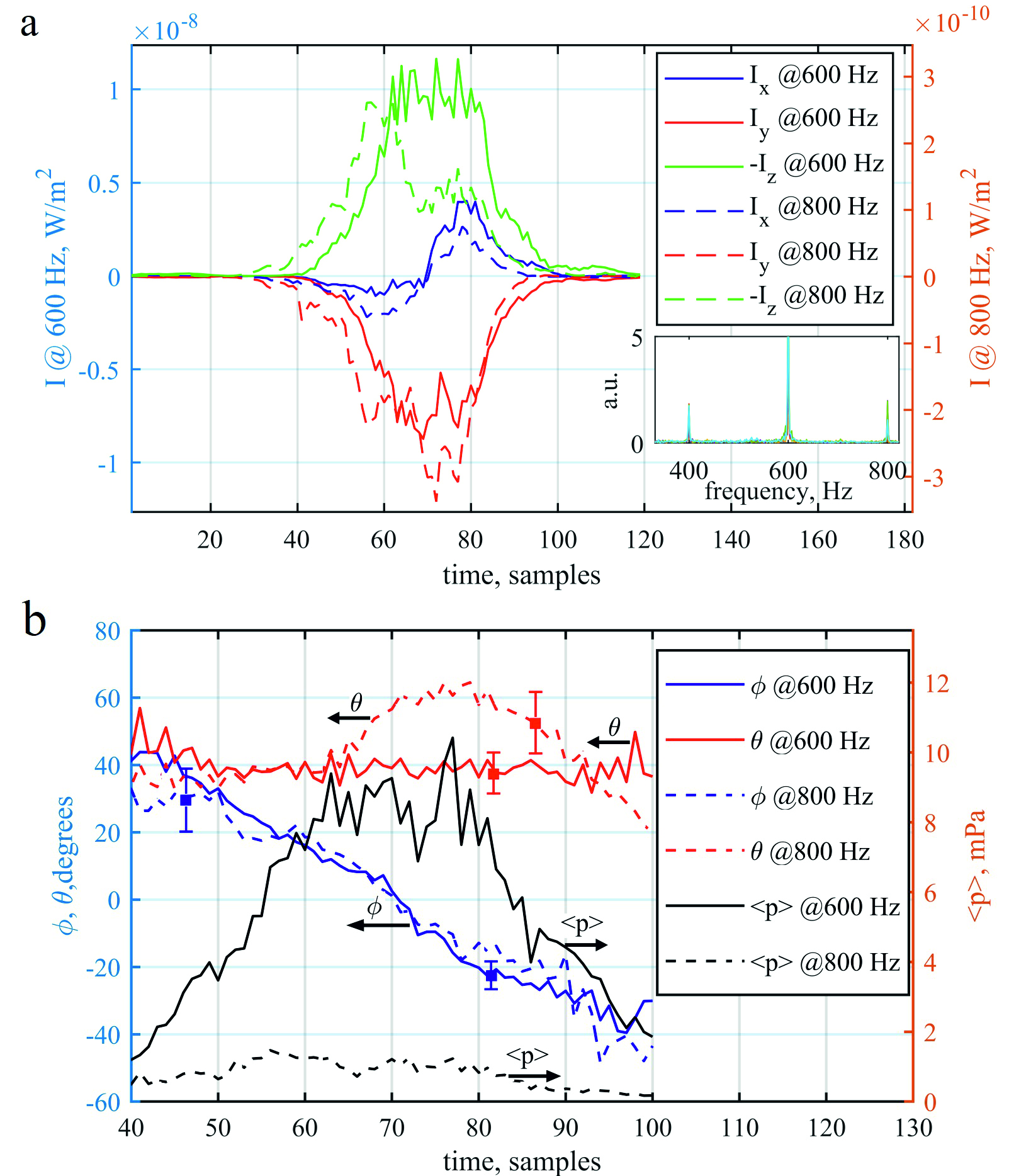}
\caption{The temporal evolution of the orthogonal components of the acoustic intensity vector (a) and the directional angles together with the average acoustic pressure (b) obtained at the frequencies 600 Hz and 800 Hz during the acoustic source movement. The inset at the upper figure shows the Fourier-spectrum of the received acoustic signal.}
\end{figure}

The bearing vector D being calculated in the same way as in the previous section was used to obtain both the azimuthal angle, $\phi$, (between D and y-axis) and the zenith angle, $\theta$, (between D and the z-axis). Fig. 4(b) shows the temporal evolution of $\phi$ and $\theta$ accompanied by evolution of the average acoustic pressure, 〈p〉, measured at both frequencies (600 and 800 Hz). As seen the level of the 〈p〉 defere for these frequencies by more than 6 folds and amount in the maximum 9 mPa and 1.5 mPa, respectively, while the background acoustic noise is at the level of 0.2 mPa. It is worth to note that, in spite of worse SNR at 800 Hz (17.5 dB) in comparison with SNR at 600 Hz (33 dB), the angles $\phi$ and $\theta$ can still be determined by means of the fiber-optic scalar-vector receiver with the RMS 20$^o$ and 3.5$^o$ obtained at the side and the central frequencies, respectively. As seen, the zenith angle, $\theta$, obtained at the frequency of 600 Hz was remained unchanged within some fluctuations during the entire measurement period. The deviation of the angle $\theta$ obtained at the frequency of 800 Hz in the period of 65\textdiv90 samples is associated with the interference of the initial acoustic wave with the wave reflected from the bottom.

\begin{figure}[htbp]
\includegraphics[width=\linewidth]{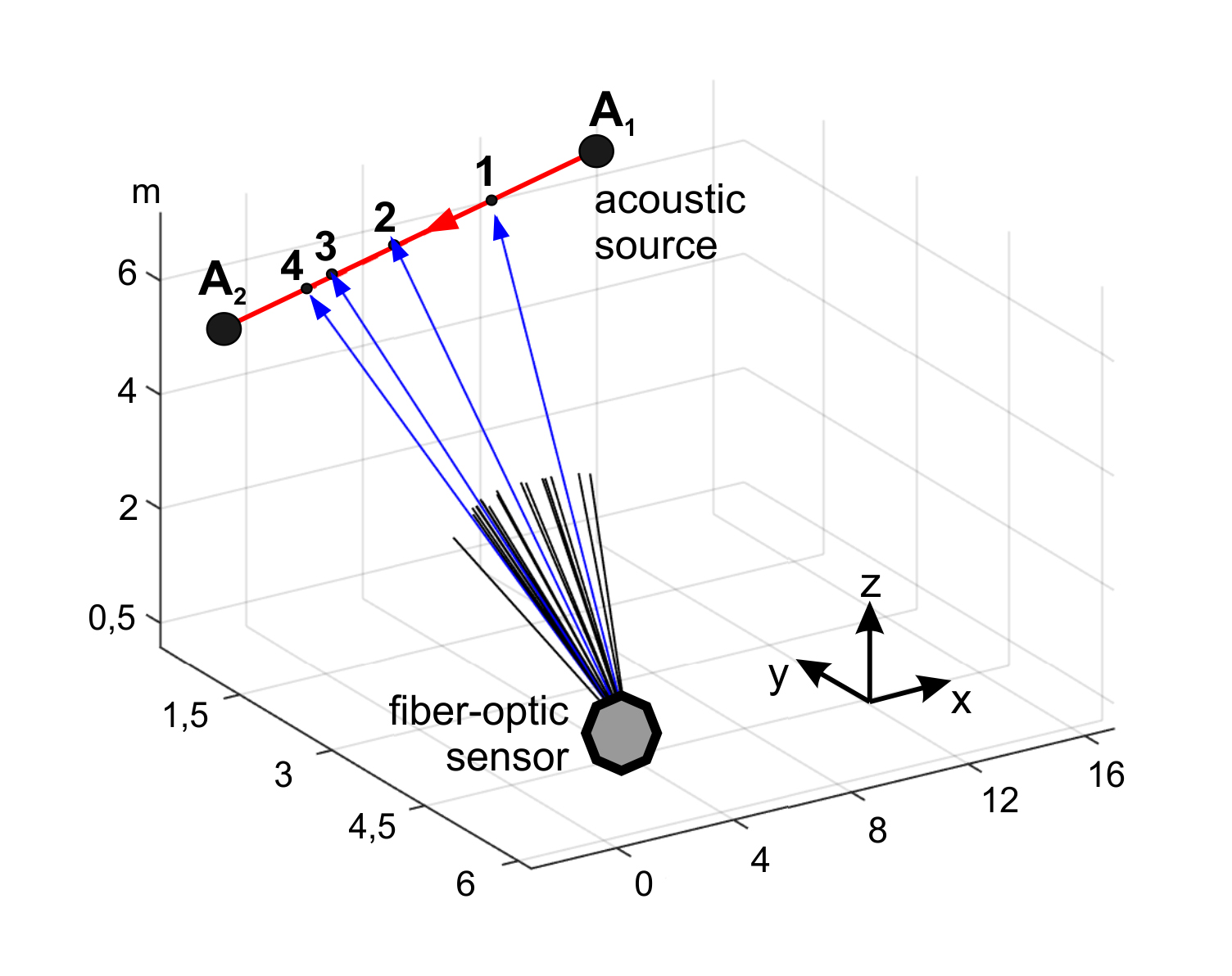}
\caption{Experimentally obtained bearing vectors (black) and their continuations (blue) to the acoustic source moving along the linear trajectory (red). Points 1, 2, 3 and 4 correspond to time moments of 40th, 60th, 70th and 75th samples, respectively.}
\end{figure}

Fig. 5 shows 3D view of the experimental space arrangement which contains the stationary fiber-optic sensor and the acoustic source moving along the trajectory A$_{1}$–A$_{2}$. The bearing vectors D obtained in the experiment at the central frequency (600 Hz) are shown by black lines. Part of them (obtained at time moments of 40th, 60th, 70th, 75th samples) are prolonged so that they intersect with the trajectory of the source movement. The intersection points coincide with the locations of the source at the corresponding time moments with appropriate accuracy.
It is clear that correctness of acoustic source bearing depends on acoustic conditions in a water area. In case of shallow sea when a reflection from water surface and bottom becomes significant, the registering the of acoustic signals at different frequencies and following application of averaging procedure can provide improvement of the bearing accuracy.

\section{Conclusion}

In this work, we developed the mobile fiber-optic scalar-vector acoustic receiver based on adaptive holographic interferometer. The receiver possesses high sensitivity to acoustic pressure at the level (11.7 $\pm$ 0.2) rad/Pa or (51 $\pm$ 0.19) V/Pa and provides a detection of weak acoustic signals with pressure above 0.5 mPa. Field tests of the developed receiver applied for obtaining scalar and vector parameters of acoustic waves generated by a stationary and moving acoustic source in open air and water area are carried out. Experimental results show perceptiveness of use of the fiber-optical adaptive interferometry system for bearing of weak acoustic sources in real conditions with accuracy not worse than 3.5$^o$. The use of the developed acoustic receiver on the base of adaptive interferometer opens up new possibilities for solving problems related to the determination of the energy, interphase, coherent and probabilistic properties of acoustic field which, in turn, can become a basis for the new effective detection algorithms and methods both for precise bearing of weak acoustic sources and their classification.

\section{Acknowledgements}

The research is supported by the Russian Science Foundation (grant No. 19-12-00323).

\providecommand{\noopsort}[1]{}\providecommand{\singleletter}[1]{#1}%

\end{document}